\documentclass[manuscript]{acmart}

\AtBeginDocument{%
  }

\setcopyright{none}

\renewcommand\footnotetextcopyrightpermission[1]{} 

\newcommand{\mycite}[1]{\citeauthor{#1}~\cite{#1}}
\usepackage{tabularx}
\usepackage{soul}
\usepackage{cleveref}
\usepackage{pgfplots}

\usepackage{tcolorbox}

\tcbuselibrary{skins}
\newtcolorbox{finding}[1]{hbox boxed title,
enhanced,attach boxed title to top left=
{yshift=-3mm,yshifttext=-1mm, xshift=1cm},
boxed title style={size=small,colback=gray!70},
title={#1}}

\begin{document}

\title{Investigating the Impact of Code Comment Inconsistency on Bug Introducing}


\author{Shiva Radmanesh}
\affiliation{%
  \institution{University of Calgary}
  \city{Calgary}
  \country{Canada}
  }
\email{shiva.radmanesh1@ucalgary.ca}

\author{Aaron Imani}
\email{aaron.imani@uci.edu}
\orcid{0000-0001-7183-5468}
\affiliation{%
  \institution{University of California, Irvine}
  \city{Irvine}
  \state{California}
  \country{USA}
}

\author{Iftekhar Ahmed}
\orcid{0000-0001-8221-5352}
\affiliation{%
  \institution{University of California, Irvine}
  \city{Irvine}
  \state{California}
  \country{USA}
}
\email{iftekha@uci.edu}

\author{Mohammad Moshirpour}
\affiliation{%
  \institution{University of California, Irvine}
  \city{Irvine}
  \state{California}
  \country{USA}
}
\email{mmoshirp@uci.edu}

\renewcommand{\shortauthors}{Radmanesh et al.}

\begin{abstract}
 Code comments are essential for clarifying code functionality, improving readability, and facilitating collaboration among developers. Despite their importance, comments often become outdated, leading to inconsistencies with the corresponding code. This can mislead developers and potentially introduce bugs. Our research investigates the impact of code-comment inconsistency on bug introduction using large language models, specifically GPT-3.5. We first compare the performance of the GPT-3.5 model with other state-of-the-art methods in detecting these inconsistencies, demonstrating the superiority of GPT-3.5 in this domain. Additionally, we analyze the temporal evolution of code-comment inconsistencies and their effect on bug proneness over various timeframes using GPT-3.5 and Odds ratio analysis. 
Our findings reveal that inconsistent changes are around 1.5 times more likely to lead to a bug-introducing commit than consistent changes, highlighting the necessity of maintaining consistent and up-to-date comments in software development.
This study provides new insights into the relationship between code-comment inconsistency and software quality, offering a comprehensive analysis of its impact over time, demonstrating that the impact of code-comment inconsistency on bug introduction is highest immediately after the inconsistency is introduced and diminishes over time.
\end{abstract}




\maketitle

\section{Introduction}
Code comments, written in natural language, are explanatory notes added by programmers within the source code of software systems and are ignored by the compiler.
They serve the purpose of clarifying the code's functionality \cite{padioleau2009listening}, aiding in code understandability, readability, \cite{kernighan1999practice, pascarella2017classifying, woodfield1981effect},  and facilitating collaboration among developers \cite{ying2005source, storey2008todo}. 
Therefore, comments play a vital role in software systems, particularly in the context of software maintenance and code updates \cite{de2005study}.

Despite their importance, comments often do not change in sync with code changes. 
This can happen due to developers forgetting or neglecting to update them \cite{zhou2017analyzing, wen2019large, ratol2017detecting, tan2007icomment}.   
As a result, comments would be inconsistent with their corresponding code which can be misleading and confusing for developers, potentially leading to future bugs \cite{tan2007icomment, fluri2007code, ibrahim2012relationship, jiang2006examining}. 
Consequently, within the research domain focused on evaluating comment quality, detecting the inconsistency between code and comment has gained significant attention due to its importance \cite{ratol2017detecting, tan2007icomment, tan2012tcomment, panthaplackel2021deep, liu2021just, liu2018automatic, 10172689, rabbi2020detecting}.

Code comments represent a crucial artifact for understanding a system, the source codes itself for maintenance purposes \cite{de2005study}. However, inconsistency between code and comments, particularly due to outdated comments, can impair software quality, potentially resulting in future bugs. In this research, we underscore the necessity of maintaining up-to-date and consistent code comments by highlighting the direct link between code-comment inconsistency and bug introduction.
Limited research has explored the effects of inconsistency between code and comments \cite{ibrahim2012relationship, tan2007icomment, khamis2010automatic}, 
their reliance on heuristic, rule-based methods for detecting inconsistencies between code and comments.

\mycite{tan2007icomment} first used Natural Language Processing (NLP) to parse comments into main clauses, sub-clauses, and semantic roles such as subjects and objects. 
Statistical techniques like clustering and correlation analysis are then used to identify hot topics and associated words within comments, enabling the tool to focus on specific topics. The predefined rule templates in their framework, structured representations of common programming rules, guide the extraction and verification process. 
Next, they employed a traditional machine learning method, decision trees, to classify new comments and map them to predefined rule templates. They utilized program analysis techniques to fill rule parameters, to identify inconsistencies between comments and code, and to rank them. 
One of the significant issues in this approach is the potential for false positives and negatives due to the complexity and variability of natural language used in comments as mentioned by them in their paper. The tool's reliance on predefined rule templates and decision tree models, while effective, may not cover all possible variations and nuances in comments, leading to inaccuracies.

\mycite{khamis2010automatic} approach for detecting inconsistencies between code and comments utilizes several heuristics to ensure comments accurately describe the corresponding code, focusing on method return types, parameters, and exceptions. It includes three key heuristics: 1) \textit{RSYNC} ensures the @return tag matches the actual return type 2) \textit{PSYNC} verifies the @param tags align with the method's parameters 3) ESYNC checks that the @throws or @exception tags accurately reflect the exceptions the method can throw. 
Similar to \mycite{tan2007icomment}, their approach is based on heuristic methods, which may not cover all possible cases of inconsistency. These heuristics, while effective in many scenarios, can miss subtler mismatches that require deeper semantic understanding.

\mycite{ibrahim2012relationship} define consistent and inconsistent changes based on whether updates to the code are accompanied by corresponding updates to the comments. They assumed that each code change should ideally be reflected in the comments, in line with established programming guidelines. Their method identified a change as inconsistent when code modifications occur without updating the associated comments, or vice versa. This approach has a limitation in not considering the semantics of the changes, potentially leading to incorrect classifications, such as when code changes may not necessitate comment updates or when simultaneous code and comment modifications do not ensure consistency, such as minor grammatical fixes.

These methods often lack the required sophistication and fail to consider the semantics and true meaning conveyed by a comment, which often leads to unrealistic results.
This may be attributed to the limited prevalence and advancement of deep learning techniques at the time, which hindered their ability to automatically capture the semantics and nuances of comments expressed in natural language.

With the rise of deep learning \cite{GoodBengCour16}, numerous researchers directed their efforts toward utilizing deep learning techniques for detecting inconsistencies between code and comments \cite{panthaplackel2021deep}, \cite{liu2021just}, \cite{louis2018deep}, \cite{liu2018automatic}, \cite{rabbi2020detecting}. 
However, there remains a gap in investigating the impact of code-comment inconsistency through more advanced and reliable methods despite these advancements in deep learning methods for detecting code-comment inconsistency.

These deep learning-based models and neural networks address the semantic and contextual meaning of comments. Following the advent of deep learning models, Large Language Models (LLMs), emerged recently. These models are designed to comprehend the complexity of natural language, encompassing multiple contexts and complicated interrelationships. 

Given that comments are written in natural human language, we became interested in the performance of LLMs in detecting code-comment inconsistencies. Considering that OpenAI GPT models are among the most widely used LLMs, we formed our initial research question:

\textbf{RQ1: How does GPT-3.5 perform in detecting inconsistency between code and comments compared to other state-of-the-art deep learning-based models?}\\
As mentioned above, inconsistency between code and comment stands out as a critical issue in software development. However, there has been limited research dedicated to examining its tangible impact on bug introduction in software systems. \mycite{tan2007icomment}  conducted a manual analysis of bug databases, revealing that 62 bug reports in FreeBSD revolved around outdated and confusing comments. they also developed a rule-based tool for automated detection of such inconsistencies, categorizing them into two types: 
1) Bugs, representing source code diverging from correct program comments 2) Bad Comment, indicating comments inconsistent with the current code. Focusing on Linux, Mozilla, Wine, and Apache, their investigation was confined to lock-related and call-related comments, identifying a total of 60 inconsistencies (33 bugs and 27 bad comments). While they manually verified reported inconsistencies, some were confirmed by the corresponding developers. 
They explored the link between code-comment inconsistencies and bugs based on a limited number of identified cases.

\mycite{khamis2010automatic} introduced an automated method for evaluating the quality of inline documentation, having consistency with the source code as one of their metrics. They used their tool to analyze modules within two open-source applications, ArgoUML and Eclipse. Notably, their observations revealed a correlation between better performance in quality assessment and fewer reported defects among the evaluated modules.
\mycite{ibrahim2012relationship} aimed to establish a correlation between comment update practices and future bugs through linear regression models predicting bug occurrences. Their study focused on three open-source systems (FreeBSD, PostgreSQL, and Eclipse). 
 The approach primarily considers syntactic changes, failing to capture the semantic correctness of comments. This means that even when comments are updated alongside code, they might still be incorrect or outdated, leading to misclassification. Thus, the analysis might overestimate the importance of small, trivial changes that do not require comment updates, marking them as inconsistent unnecessarily.


To address limitations and deficiencies identified in previous research, we conducted a detailed analysis of the changes associated with bug-introducing and non-bug-introducing commits, utilizing a high-performance model for code-comment inconsistency detection (GPT-3.5). We specifically focus on examining the history of changes that occurred within specific time frames preceding the occurrence of these (non)bug-introducing commits.

This prompted the formulation of our first research question:\\
\textbf{RQ2: Does inconsistency between comments and code impact bug proneness?}\\
Given the continual evolution of software systems over time, investigating the timing when outdated comments are more likely to introduce bugs is crucial for understanding their impact on bug-introducing in software development processes.
Previous studies have touched upon the impact of code comment inconsistency on bugs, yet they have encountered accuracy issues and methodological limitations \cite{tan2007icomment}, \cite{khamis2010automatic}, \cite{ibrahim2012relationship}. However, there exists a gap in understanding the evolving impact of code comment inconsistency on bug-proneness over time.
In this research, we aim to address this gap by conducting a large-scale analysis of the impact of code comment inconsistency on bug introduction within various timeframes leading up to the occurrence of commits that either introduce or do not introduce bugs. This investigation will allow us to answer our next research question:
\\
\textbf{RQ3: How does the impact of outdated comments on bug-proneness change over time?}

After exploring the impact of code-comment inconsistency on bug introduction and its evolution over time, we aim to compare our approach and findings with previous studies \cite{tan2007icomment, khamis2010automatic, ibrahim2012relationship}.
To answer our research questions, we first used a publicly available dataset for code comment inconsistency detection. After fine-tuning the GPT-3.5 model on a limited sample from the training and validation sets, we tasked it with detecting inconsistencies for each test set record. This allowed us to compare GPT-3.5's results with those of other state-of-the-art deep learning models, revealing GPT-3.5's superior performance, even with minimal fine-tuning (RQ1).
To investigate the impact of code-comment inconsistency on bug-introducing, we conducted a large-scale analysis using Apache projects due to their well-documented nature and extensive historical data. For each Apache repository, we employed GitCProc \cite{casalnuovo2017gitcproc} to identify bug-fix commits. Subsequently, we utilized the SZZ algorithm\cite{rosa2021evaluating, rosa2023comprehensive} with the input of these bug-fix commits identified by GitCProc.
We analyzed a sample of $N$ bug-introducing commits and to have a baseline, we did the same analysis for the same number of non-bug-introducing commits in each repository, comparing them with all previous commits within the last 7 days of occurrence. Subsequently, we identified methods in previous commits differing from their counterparts in the sampled (non)bug-introducing commits. From each identified method, we created a data record containing the old and the new method and their comments. 
For each data record, we instructed GPT 3.5 to compare the new comment with both the new code and the old code to identify data records that contain changes with outdated comments, as RQ1 demonstrated GPT-3.5's superiority in detecting code-comment inconsistencies. Then, through statistical analysis and odds ratio analysis, we showed how commits with detected inconsistent changes within the past 7 days of their commit date are likely to be bug-introducing compared to those without inconsistencies in that time window (RQ2).
In order to understand the evolution of the impact of code comment inconsistency on bug-introducing, we conducted a similar analysis using a 14-day timeframe. We then compared the ratio of the impact of code comment inconsistency on bug-introducing in each timeframe (RQ3).
\\
In summary, our contributions are as follows:
\begin{itemize}
    \item We compare state-of-the-art deep learning-based models with one of the most commonly used LLMs for detecting code comment inconsistency.
    \item We demonstrate the impact of code comment inconsistency on bug-proneness, and do a comparative analysis of our approach and findings with existing literature, uncovering novel insights and contributions not previously explored.
    \item We analyze the evolution of the impact of code comment consistency on bug-proneness over time.
\end{itemize}
Our implementations are provided in the supplementary materials \cite{a2024_anonymous}.

\section{Related Work}

Code comments are crucial for enhancing readability and maintaining software. Prior studies have focused on comment analysis, code comment quality evaluation tools, code comment update and generation tools, and the impact of outdated comments. Our research is most closely aligned with the latter.This section reviews the literature on these topics, highlighting the gaps our study addresses and our contributions to the area.

\subsection{Code Comments Analysis}
An experiment was conducted by \mycite{woodfield1981effect} to assess the impact of modularization types and comments on programmers' comprehension of programs. They found that including comments in the program significantly enhances programmers' comprehension.

\mycite{ying2005source} investigated the role of code comments within the IBM internal codebase and found out that programmers utilize comments not only for explaining code but also for various other functions, including team communication.

\mycite{jiang2006examining} analyzed the evolution of code comments in the PostgreSQL project. Their study revealed that the percentage of functions with header and non-header comments remained stable over time, except for initial fluctuations caused by the introduction of new commenting styles.

\mycite{fluri2007code} presented an approach to track the co-evolution of code and comments across multiple versions. Their analysis of three open-source systems (ArgoUML, Azureus, and JDT Core) showed that newly added code is scarcely commented, and a significant 97\% of comment changes occur in the same revision as the corresponding source code change.
In their following study, \mycite{fluri2009analyzing} introduced an approach for associating comments with source code entities to track their co-evolution across multiple versions. The study analyzed the co-evolution of code and comments in eight diverse software systems and revealed that the proportion of comments to source code increases similarly across all eight systems, but newly added code was not consistently well-commented. The likelihood of a source code entity receiving comments depends on its type, and In six out of eight systems that they analyzed, comments and their associated source code changes co-evolve in more than 90\% of all comment changes.

\mycite{arafat2009commenting} explored the development processes of 5,229 open-source projects, focusing on the density of comments in open-source code as a quality indicator. They found that successful open-source projects are consistently well-documented with an average comment density of 18.67\% and this comment density remains stable regardless of team or project size.

\mycite{pascarella2017classifying} investigated different purposes of code comments in software development. In their study, they examined six Java open-source projects,  to understand the varied purposes of code comments. Through this analysis, they introduced a hierarchical taxonomy of source code comments, comprising 16 inner categories and 6 top categories, and manually classified over 2,000 code comments from the mentioned projects. 

\mycite{wen2019large}  investigated how code and comments co-evolve by mining 1.3 Billion AST-level changes from the complete history of 1,500 systems. They also manually analyzed 500 commits to define a taxonomy of code-comment inconsistencies fixed by developers.
This analysis revealed the extent to which various types of code changes prompt updates to associated comments, highlighting scenarios where code-comment inconsistencies are more likely to occur. 

\mycite{rani2023decade} conducted a systematic literature review on code comment quality assessment.

\subsection{Code Comment Quality Evaluation Tools}
The problem of assessing code comment quality has gained a lot of attention among researchers. In this area, there has been a significant emphasis on detecting inconsistencies between code and comments in many studies.  \mycite{khamis2010automatic} introduced JavadocMiner, an automated approach for evaluating the quality of inline documentation, using a set of heuristics to assess both the quality of language and consistency between source code and its comments. 
\mycite{steidl2013quality} presented a detailed approach for quality analysis and evaluation of code comments. They used machine learning on Java and C/C++ programs, and categorized comments based on their types. They provided a model for comment quality that describes quality attributes, namely coherence, consistency, completeness, and usefulness. 

Several rule-based methodologies have been suggested to detect inconsistencies between code and comments. \mycite{tan2007icomment} introduced iComment, a tool to automatically analyze comments and detect comment-code inconsistencies. iComment extracts implicit program rules from comments and leverages them to detect inconsistencies with the code by using techniques from Natural Language Processing (NLP), Machine Learning, Statistics, and Program Analysis. In their following study \mycite{tan2012tcomment} introduced @TCOMMENT, an approach for testing the consistency of Java method properties described in Javadoc comments, particularly those related to null values and exceptions.
\mycite{ratol2017detecting} designed a rule-based approach, called FRACO, to detect inconsistent comments with respect to rename refactorings by considering factors like identifier type, morphology, scope, and comment location. FRACO outperformed Eclipse’s automated in-comment identifier replacement feature.
\mycite{zhou2017analyzing} presented an automated approach to detect defects in API documents, specifically focusing on directives related to parameter constraints and exception-throwing declarations. They leveraged techniques from program comprehension and natural language processing. Their approach used a first-order logic-based constraint solver to identify to detect the defects in case of inconsistency.

More recently, machine learning and deep learning have found extensive applications in outdated comment detection. For example, \mycite{liu2018automatic} introduced a machine learning-based method for detecting outdated comments during code changes by leveraging 64 features derived from code changes, comments, and analyzing the relationship between code and comments before and after modifications. 
\mycite{rabbi2020detecting} introduced an approach for detecting code comment inconsistencies, using a siamese recurrent network that considers both word tokens and their sequences in codes and comments. Their network has 2 separate RNN-LSTM models to analyze codes and comments, generate vectors for comparison, and predict inconsistencies by measuring the similarity between these vectors.

Subsequently, various methods were introduced for just-in-time detection of outdated comments alongside code changes.
\mycite{panthaplackel2021deep}  introduced a deep learning approach for just-in-time detection of inconsistency between comments and code by training a model to correlate comments with code changes. 
\mycite{liu2021just} developed a system for just-in-time obsolete comment detection and update, which includes an obsolete comment detector (OCD) and a comment updater (CUP) each employing distinct neural network models for detection and updates, respectively.
Recently \mycite{10172689} introduced ADVOC, an approach for obsolete comment detection with a key emphasis on training data quality. ADVOC includes a data cleaning module for rectifying false positive samples, a data sample encoding module for capturing complex semantics between code changes and comments, and a detection model learning module that employs adversarial learning to handle false negative samples.

\subsection{Code Comment Update and Generation Tools}
Researchers have widely explored techniques for comment generation and updating to enhance the overall quality of code comments and update outdated comments.
\mycite{sridhara2010towards} introduced a technique to automatically generate comprehensive summary comments for Java methods using the structural and linguistic cues within the method.
\mycite{haiduc2010use} explored various text summarization techniques like Text Rank for generating source code summaries. 
\mycite{moreno2013automatic} introduced a heuristic-based technique to automatically generate human-readable summaries for Java classes using information about the class stereotypes.
\mycite{subramanian2014live} introduced Baker, an iterative deductive approach for linking source code examples with API documentation. Baker supports both Java and JavaScript.
\mycite{mcburney2014automatic} introduced an approach for automatically generating summaries of Java methods, focusing on providing context surrounding a method rather than internal details by leveraging  PageRank to identify crucial methods in the context, and SWUM  to gather relevant keywords describing their behavior.
\mycite{dagenais2014using} introduced AdDoc, a technique designed to automatically identify and track documentation patterns within code and their documentation, which can help contributors to adapt existing documentation.
\mycite{wong2015clocom} proposed CloCom, an approach to automatically generate code comments by analyzing existing software repositories.  It utilizes code clone detection techniques to find similar code segments and uses the comments from some segments to describe others.  Using Natural language processing techniques, they select relevant comment sentences.
These studies primarily utilized rule-based and heuristic-based methods.

However, more recent research studies have employed machine learning and deep learning techniques for code comment generation and update.
\mycite{iyer2016summarizing} introduced CODE-NN, an end-to-end neural attention model using Long Short Term Memory (LSTM) networks to produce descriptive sentences for C\# code snippets and SQL queries.
\mycite{allamanis2016convolutional} proposed an attentional neural network architecture for the summarization of source code snippets. The architecture utilizes two attention mechanisms to generate the summary: one predicts the next token based on attention weights, and the other can copy a code token directly into the summary.
\mycite{hu2018deep} introduced DeepCom, an attention-based Seq2Seq model for automatically generating code comments for Java methods. DeepCom takes AST sequences as input and converts them to specially formatted sequences using a new structure-based traversal (SBT) method, which can express the structural information and keep the representation lossless at the same time.
In a follow-up research \mycite{hu2020deep} propose Hybrid-DeepCom which is an extended version of DeepCom.  It combines the source code and the AST sequences together to generate comments by leveraging both lexical and syntactical information.
\mycite{leclair2019neural} introduced a neural model called ast-attendgru, that combines information from code words and code structure from an Abstract Syntax Tree (AST), which allows the model to learn code structure independently of the text in the code, enabling it to generate coherent summaries for source code even in cases where there is little or no internal documentation. 
\mycite{zhou2019augmenting} introduced ContextCC which leverages a Seq2Seq Neural Network model with an attention mechanism to automatically generate concise comments for Java methods.
\mycite{liu2021just}  introduced CUP2, a two-stage framework designed to address the issue of obsolete comments in source code. It comprises two stages: an Obsolete Comment Detector (OCD) and a Comment Updater (CUP), both leveraging distinct neural network models for detection and updates, respectively. OCD is responsible for determining whether a given comment should be updated, while CUP generates the updated comment if necessary. 
\mycite{lin2021automated} proposed HEBCUP, a heuristic-based approach for code comment updates, which is designed based on the observations of CUP's successful and failure cases and focuses on updating comments with changed code.
\mycite{zhu2022hatcup} introduced HatCUP which uses an RNN-based encoder-decoder architecture. HatCUP considers code structure change information, utilizing a structure-guided attention mechanism, code change graph analysis, and data flow dependency analysis.
\mycite{shi2022evaluation} conducted a comprehensive analysis of recent neural code summarization models, with a focus on evaluating various aspects that impact the effectiveness of these models.

\subsection{Impact of Code-Comment Inconsistency}

Very few research works have investigated the impact of
code comment inconsistency. \mycite{tan2007icomment} evaluated their tool, iComment, designed for automated detection of inconsistencies between comments and source code, on four large code bases: Linux, Mozilla,
Wine, and Apache to find outdated comments or bugs (where source code does not follow the documentation). Many of the code-comment inconsistencies detected by iComment were confirmed by the developers based according to their study, and some developers acknowledged that these inconsistencies had the potential to mislead developers and introduce future bugs. Additionally, the authors presented two instances of outdated comments they detected on Mozilla that had caused a new bug reported in later versions of Mozilla.
\mycite{fluri2007code} investigated the practices of updating code comments and their impact on software reliability. It highlighted that inconsistent updates between code and comments can lead to outdated comments, potentially causing future bugs. Their study focused on three large open-source systems (FreeBSD, PostgreSQL, and Eclipse) and established a connection between comment update practices and future bugs by using linear regression models that predict the number of future bugs.
\mycite{khamis2010automatic} presented an automated method to assess the quality of inline documentation, focusing on both language quality and consistency with the source code.  they applied their tool to the modules of two open-source applications (ArgoUML and Eclipse). They correlated the results returned by the analysis with bug defects reported for the individual modules.

\section{Methodology}

In this section, we first evaluate the performance of the GPT-3.5 model in detecting inconsistencies between code and comments, comparing its effectiveness with other contemporary deep learning-based models (RQ1), then we explain our methodology for assessing the impact of code-comment inconsistency on bug-introducing using GPT-3.5 (RQ2) across different timeframes preceding commits that either introduce or do not introduce bugs (RQ3), and finally doing a comparison among previous approaches and ours for analyzing the impact of code comment inconsistency on bug-introducing.

\subsection{Evaluating GPT-3.5 against other models}
 Following, we detail our approach in assessing the potential of GPT-3.5 in detecting code comment inconsistency (RQ1).

\subsubsection*{\textbf{Dataset}}
To evaluate the performance of the GPT-3.5 in comparison to other deep learning models for detecting inconsistencies between code and comments, we conducted an analysis using the cleaned version of the CUP2 \cite{liu2021just} dataset provided by \mycite{10172689}. The original version of this dataset was introduced by \mycite{liu2021just}.
After cleaning and enhancing the CUP2 dataset, \mycite{10172689} randomly selected 1000 instances and manually corrected their labels to create the "verified test set". This subset includes 500 negative and 500 positive labels, making it a balanced set. While the full-test quality was also improved, only the labels in the verified test set were manually checked and confirmed. We utilized both the full test set and the verified test set of this refined dataset to evaluate the performance of the GPT-3.5 against other baseline and state-of-the-art deep learning models in detecting inconsistencies between code and comments.

\subsubsection*{\textbf{Evaluation}}
We randomly sampled a subset of the train and validation set and used them as a validation set for our prompt engineering. Subsequently, we applied the prompt that exhibited the highest performance on this set to the entire test set. We then compared the outputs of GPT-3.5 with baselines in detecting code-comment inconsistencies, namely, Fraco\cite{ratol2017detecting}, Random Forest algorithm, OCD (the detector model of the CUP2)\cite{10172689}, the just-in-time inconsistency detection between comments and source code model introduced by \mycite{panthaplackel2021deep}, and ADVOC\cite{10172689} on this dataset. The evaluation metrics used were precision, recall, and F1-score.

\subsubsection*{\textbf{Prompt Engineering}}
In this section, we discuss the prompt engineering steps we took in optimizing the GPT-3.5 for the best performance performance. 
In addition to employing prompt engineering strategies provided by OpenAI, we adopted a combination of multiple prompt patterns from a catalog of successful patterns for enhancing LLM conversations by \mycite{white2023prompt}, to achieve optimal results. This catalog is widely utilized by researchers for prompt engineering\cite{zhao2023survey, kocon2023chatgpt, white2024chatgpt}.

In our initial step, we utilized the \textit{``Question Refinement Pattern''} from the mentioned pattern catalog\cite{white2023prompt} to improve the design and wording of our general prompt. This pattern involves engaging the LLM in prompt engineering by explaining the task and requesting a better version of the question. As a result, the input structure became more detailed, and the task definition clearer.
Additionally, we employed the \textit{``Template Pattern''} to ensure that the output of GPT-3.5 follows a specific template structure. This was necessary for automated analysis of the responses. We instructed the LLM to use a JSON format for the response.
Furthermore, we also explored hyperparameter tuning by optimizing temperature settings and p\_sampling thresholds to enhance response diversity and relevance.


\subsubsection{Prompting the LLM}
\paragraph{Zero-Shot}
Initially, we investigated a zero-shot prompting approach\cite{liu2023pre} followed by prompt engineering. This method aimed to leverage pre-trained language models without additional task-specific training.

\paragraph{Few-Shot}
Then, we experimented with few-shot prompting \cite{liu2023pre} by providing the model with a few examples in the prompt to improve its accuracy. This method aimed to enhance performance through contextual learning from limited data points.

\paragraph{Finetuning}
The potential of GPT-3.5 extends far beyond the zero-shot method. Its adaptability to diverse tasks enables adaptation to diverse tasks without extensive fine-tuning.
For the task of code-comment inconsistency detection, our approach involved fine-tuning the GPT-3.5 using a limited set of code-comment inconsistency detection samples. Through this process, the model rapidly acquires task-specific knowledge based on a small set of examples. Consequently, the model learns to generalize effectively for the new task.

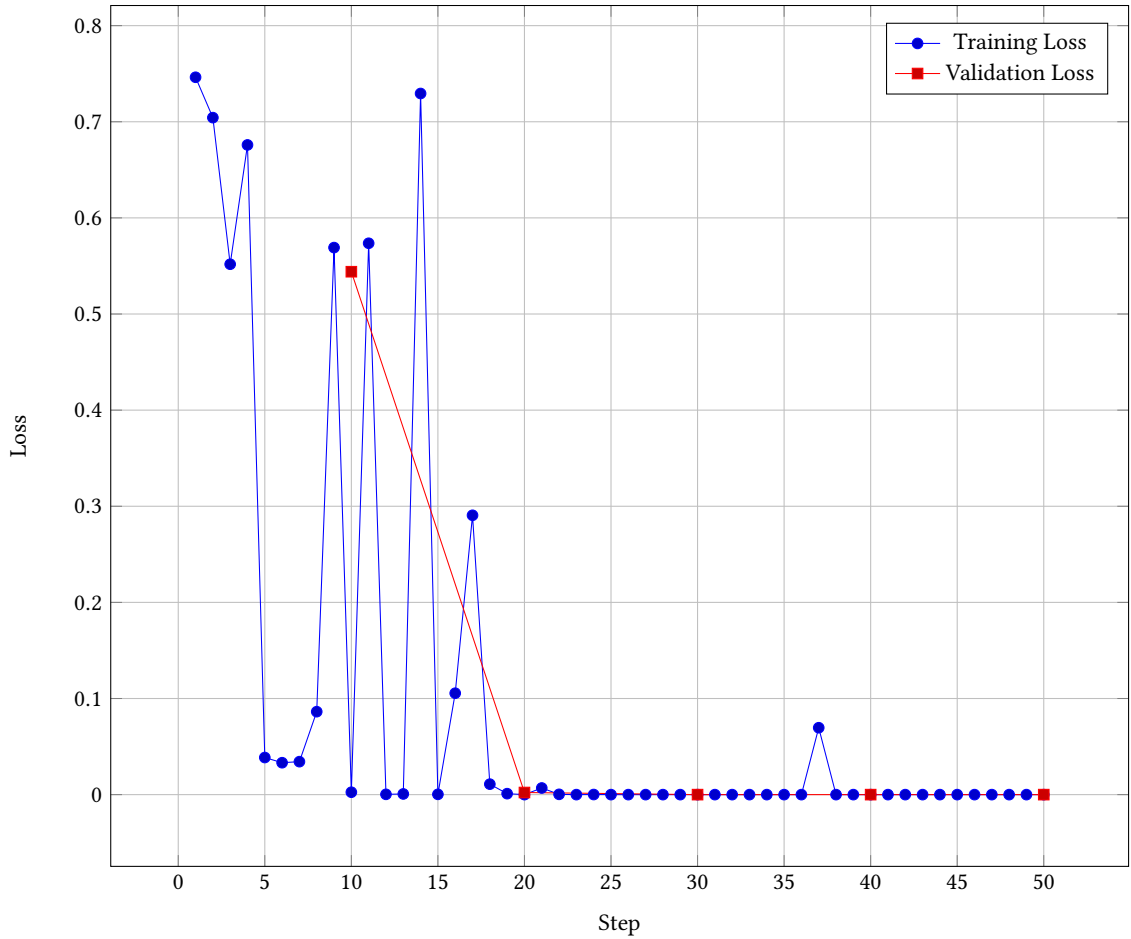
\begin{figure}
    \centering
\begin{tikzpicture}
    \begin{axis}[
        xlabel={Step},
        ylabel={Loss},
        title={},
        grid=major,
        width=\textwidth,
        unbounded coords=discard, 
    ]
    \addplot table [
        col sep=comma,
        x=Step,
        y=Training loss,
        restrict expr to domain={\thisrow{Step}}{0:50}
    ] {files/train-val-loss.csv};
    \addplot table [
        col sep=comma,
        x=Step,
        y=Validation loss,
        restrict expr to domain={\thisrow{Step}}{0:50},
                unbounded coords=discard, 
        ] {files/train-val-loos.csv};
    \legend{Training Loss, Validation Loss}
    \end{axis}
\end{tikzpicture}
    \caption{Training and Validation Loss Over Steps}
    \Description[Drop in Validation Loss]{The plot illustrates that the validation loss decreases to zero after 20 iterations.}
    \label{fig:train-val-loss}
\end{figure}

To accomplish this objective, we extracted a subset from the training and validation sets of the CUP2 dataset. Specifically, based on OpenAI's recommendation, we selected 50 training samples, comprising 25 positive and 25 negative instances, and 20 validation samples, containing 10 positive and 10 negative instances. These samples were randomly chosen and manually verified to ensure a broad spectrum of topics from various projects. Additionally, we manually verified the accuracy of the labels in the original dataset. The training loss details are shown in \Cref{fig:train-val-loss}.

\subsection{Impact of Code-Comment Inconsistency}\label{impact}
\subsubsection{Dataset}
In existing datasets provided by researchers for inconsistency detection between code and comments, instances are structured as <old code, new code, old comment, new comment>. The labeling of each instance is based on the obsoletion of the old comments \cite{panthaplackel2021deep, liu2021just, 10172689}; 
If the new comment differs (after punctuation removal) from the old comment, the instance receives a positive label, indicating inconsistency between the old comment and the new code. Otherwise, the label is negative.
\\
However, this labeling methodology presents challenges for our analysis. Given our focus on identifying inconsistencies between code and comments that could potentially lead to future bugs, we must assess the consistency between the new comment and the new code, rather than solely relying on the comparison between the old comment and the new code.
For example, consider a scenario where during a code change, a method and its comment are updated such that the new comment accurately describes the code, and the old comment accurately describes the old code as well. With those approaches, this entry would be a positive record, since the old comment and new comment differ. However, we cannot consider this change as a potential reason for future bugs because the code and its comment in the project have always been consistent.
Therefore, we developed a dataset focusing specifically on inconsistencies between the new code and its accompanying comment. This ensures that our analysis targets potential sources of future bugs within the code.

\paragraph{Project Selection}
We focused on Java projects as Java stands as one of the most popular programming languages in use today.
Specifically, we investigated 32 Apache projects written in Java due to their extensive use among researchers \cite{ahmed2015empirical, li2023commit, mannan2020relationship, chen2017characterizing, gharehyazie2015developer, kabinna2016logging}.
These projects are known for their systematic approach in maintenance and structured contribution guidelines, increasing the likelihood of comprehensive and well-structured comments throughout the codebase.

\begin{figure*}[t]
    \centering
    \includegraphics[width=\textwidth]{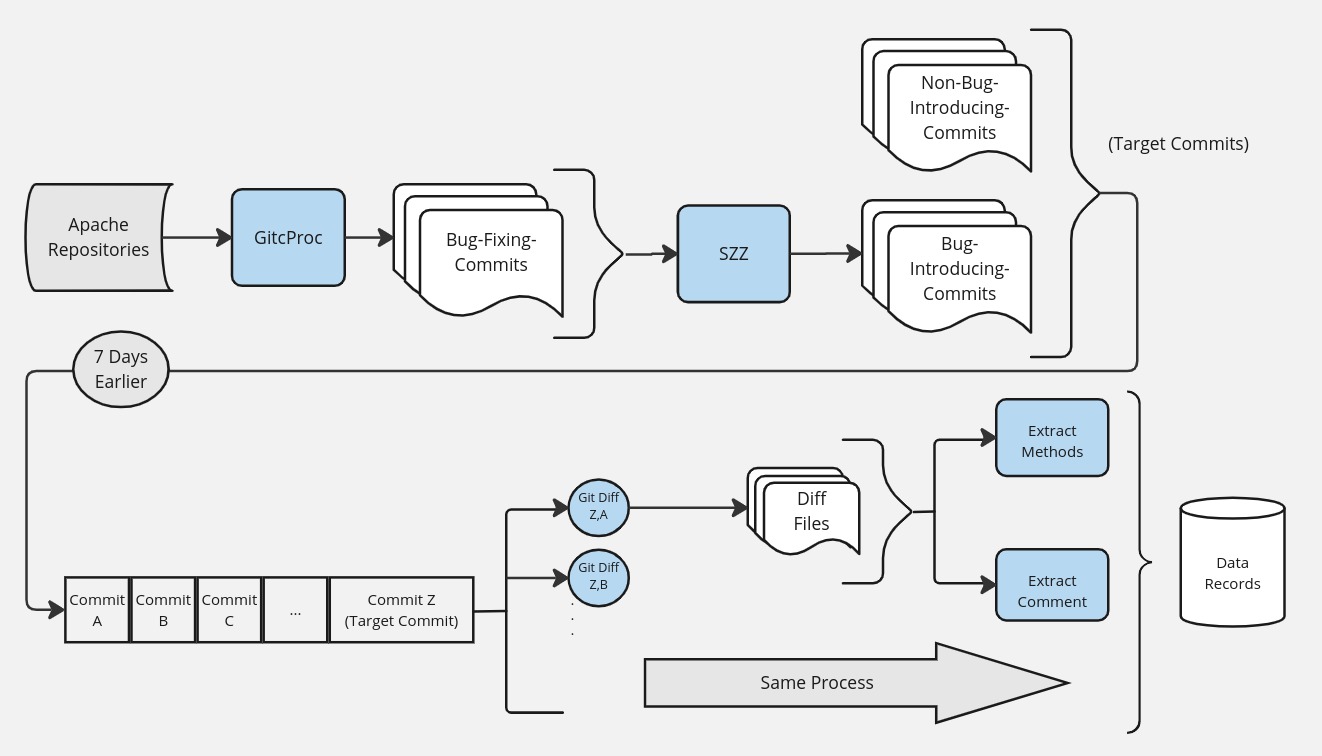}
    \caption{Our pipeline for collecting data for RQ2 and RQ3}
    \Description[data pipeline flow]{The diagram illustrates the data collection pipeline flow, as described in section 3.2}
    \label{fig:datapipline}
\end{figure*}

\paragraph{Data Curation}  
For each Apache project repository, we utilized GitCProc \cite{casalnuovo2017gitcproc}, a tool designed to identify bug-fixing commits. This enabled us to identify bug-fixing commits within each repository. Additionally, we employed the SZZ algorithm \cite{rosa2021evaluating, rosa2023comprehensive}, an approach for identifying bug-introducing commits based on bug-fixing commits in a software repository. These tools enabled us to find bug-introducing commits within each repository. 
We selected a random sample of bug-introducing commits from each repository, ensuring a confidence level of 90\% and a margin of error of 10\%. To establish a baseline, an equivalent number of non-bug-introducing commits were randomly sampled from each repository. We then filtered out those non-bug-introducing commits that only affected documentation or tests, using the remaining as our non-bug-introducing set for our experiment.
We refer to our sample of bug-introducing and non-bug-introducing commits in each repository as ``target commits''.
In each repository, we compared each target commit with all previous commits within the last 7 days of their occurrence using git diff logs. From the Git diff log, we identified the modified Java files. Within each altered file, we analyzed the old and new versions of the syntax tree to identify common methods and extract their contents. Additionally, we stored the differing methods between the old and new commits. Summary comments located above method declarations for each modified method were also extracted. These method-comment pairs form the basis of our data records, which will be elaborated on in the subsequent section.
Lastly, we excluded data records with empty comments. This decision aligns with the research goal of investigating the impact of outdated comments.

\paragraph{Dataset Structure}

The structure of each data record in our dataset includes the following components: old commit hash, new commit hash (from the target commits), old and new code snippets, old and new comment, a variable indicating whether the new commit introduces a bug or not, and additional metadata such as the dates of the old and new commits, repository name, etc.

\subsubsection{Impact Analysis}

In investigating the potential impact of code-comment inconsistency on bug introduction, the initial step involves detecting such inconsistencies within data records.
 Leveraging the LLM model, specifically GPT-3.5, we designed a prompt to assess inconsistency within each record. This assessment focused on comparing the new code with both the new and old comments. 
 Due to high cost concerns associated with using LLMs, we limited our analysis of the impact of code-comment inconsistency on bug-introducing to 8 out of the 32 Apache repositories from which we created data records. Our analysis categorized data records into three distinct groups:
\begin{enumerate}
    \item \textbf{Outdated records}, where the new comment is inconsistent with the new code but consistent with the old code. This indicates negligence in updating the comment during code modifications, resulting in inconsistency that has happened within our timeframe and existing at the time of the new commit.
    \item \textbf{Earlier outdated records}, characterized by the inconsistency between the new comment and both the old and new code, where the new comment remains unchanged from the old comment. In such records, inconsistency had been introduced before the analyzed timeframe, preceding the occurrence of the new commit.
    \item \textbf{Normal records},  where the new code is consistent with the new comment, indicating consistency during the new commit's occurrence. These records did not show any inconsistency issues.
\end{enumerate}

\paragraph{Prompt Engineering}
For the prompt engineering task in this experiment, we closely followed the methodology described in the RQ1, with an additional step incorporated. To address potential issues such as incomplete, incorrect, or ambiguous answers from LLM, we applied the \textit{``Reflection Pattern''} \mycite{white2023prompt}. This involves prompting the LLM to automatically explain the rationale behind its answer, ensuring the provided information's accuracy and alignment with our expectations. Given the absence of ground truth labels for our data records, this step enabled us to manually evaluate the quality of consistency detection in our prompts by analyzing the LLM's rationale.  This step was not necessary for the RQ1 approach, as the utilized dataset provided labels for each record, serving as ground truth for evaluating LLM's performance with the given prompt.

\paragraph{Odds Ratio Analysis}
An odds ratio quantifies the association between variables and the probability of an event. Specifically, it measures how probable an exposure is to result in a particular event. It indicates the likelihood that an exposure leads to a specific event\cite{tenny2017odds}. The odds ratio is calculated as the ratio of the odds of exposure in the group where the event occurs to the odds of exposure in the group where the event does not occur. An odds ratio greater than one shows higher odds of the event with exposure, while an odds ratio less than one shows reduced odds of the event with exposure.

In our study, the exposed group comprises Outdated and Earlier Outdated data records, while the non-exposed group consists of Normal data records. The event of interest is whether these records lead to a bug-introducing commit, i.e., whether their new commit is identified as a bug-introducing commit.
In the group where the event occurs, the odds are calculated as the ratio of outdated and earlier outdated data records leading to bug-introducing commits to normal records leading to bug-introducing commits. Conversely, in the group where the event does not occur, the odds are computed as the ratio of outdated and earlier outdated data records not leading to bug-introducing commits to normal records not leading to bug-introducing commits.

We repeated the experiment with a 14-day timeframe to investigate the impact of code comment inconsistency on bug introduction across different timeframes (RQ3).
Detailed results of the odds ratio calculation using the described setup for 7-day and 14-day windows are presented in \Cref{results}.

\section{Results}\label{results}

\subsection{Evaluating GPT-3.5 against other models}

To address RQ1, we employed the CUP2 dataset \cite{liu2021just} for detecting code comment inconsistencies, initially introduced by \mycite{liu2021just} and provided by \mycite{10172689} after cleaning. Instances with a positive label indicate inconsistency between old comments and new code. The total test set comprises 385,258 instances, with 7,359 labeled as positive (inconsistent). To ensure reliability, a verified test set of 1,000 instances (500 positives, 500 negatives) was manually checked by \mycite{10172689}. 
Our analysis involved prompt engineering and three experimental approaches: zero-shot, few-shot, and fine-tuning. The results, detailed in Table \ref{comaprison_gpt}, indicate that the fine-tuning approach outperformed the other two.

\begin{table}[b]
   \begin{tabular}{l c c c}
     &\textbf{Precision} & \textbf{Recall} & \textbf{F1-score}\\
     \toprule
    \textbf{Zero-shot}& $68.56$\% & $45.8$\% & $54.91$\%\\
    \textbf{Few-shot}& $75.80$\% & $52$\% & $61.68$\%\\
    \textbf{Fine-tuning}& $94.4$\% & $88.30$\% & $82.95$\%\\
     \bottomrule
   \end{tabular} 
   \caption{Comparison of the LLM performance under different prompting techniques}
   \label{comaprison_gpt}
\end{table}

We then evaluated the performance of our fine-tuned GPT-3.5 model on both the total test set and the verified sample, comparing it with state-of-the-art models reported by \mycite{10172689}. These models \cite{liu2021just, 10172689, panthaplackel2021deep}, outperform earlier discussed rule-based and heuristic-based methods used in \cite{ibrahim2012relationship}, \cite{khamis2010automatic}, \cite{tan2007icomment} for detecting inconsistencies between code and comments. The results of these comparisons are demonstrated in Table \ref{comparison_vgt} and Table \ref{comparison-gt}. The results from tools other than GPT-3.5 are directly sourced from \cite{10172689}.

\begin{table}[b]
   \begin{tabular}{l c c c}
     &\textbf{Precision} & \textbf{Recall} & \textbf{F1-score}\\
     \toprule
    \textbf{Fraco}& $19.5$\% & $15.4$\% & $17.2$\%\\
    \textbf{RandomForest}& $65.1$\% & $13.5$\% & $22.4$\%\\
    \textbf{OCD}& $60.2$\% & $21.6$\% & $31.8$\%\\
    \textbf{Just-In-Time} & $60.6$\% & $35.4$\% & $44.7$ \%\\
    \textbf{ADVOC} &  $60.6$\% & $35.4$\% & $44.7$\%\\
    \textbf{GPT-3.5 (fine-tuned)} & $90.97$\% & $96.38$\% & $93.60$\%\\
     \bottomrule
   \end{tabular} 
   \caption{Comparison results on the full test set of the cleaned dataset}
   \label{comparison_vgt}
\end{table}

\begin{table}[h!]
   \begin{tabular}{l c c c}
     &\textbf{Precision} & \textbf{Recall} & \textbf{F1-score}\\
     \toprule
    \textbf{Fraco}& $90.2$\% & $22.0$\% & $35.4$\%\\
    \textbf{RandomForest}& $99.1$\% & $21.4$\% & $35.2$\%\\
    \textbf{OCD}& $95.8$\% & $31.6$\% & $47.5$\%\\
    \textbf{Just-In-Time} & $97.0$\% & $51.2$\% & $67.0$ \%\\
    \textbf{ADVOC}& $98.7$\% & $58.8$\% & $73.6$\%\\
    \textbf{GPT-3.5 (fine-tuned)}&  $94.4$\% & $82.95$\% & $88.30$\%\\
     \bottomrule
   \end{tabular}
   \caption{Comparison results on the verified test sample of the cleaned dataset}
   \label{comparison-gt}
\end{table}

\subsection{Impact of Code-Comment Inconsistency}
Our approach involved analyzing N bug-introducing and non-bug-introducing commits in each repository, focusing on new commits. We compared each of these N commits with all previous commits within the last 7 days of their occurrence. Next, we searched for methods in previous commits that differed from their counterparts in the new commits. From each identified method, we created a data record containing the old method and comment from the previous commit, along with the new method and comment from one of the N new commits.

In each data record, we assess whether the new comment is consistent with the new code and with the old code. Inconsistency between the new comment and the new code signifies a present inconsistency at the time of the new commit. If the new comment matches the old code but is inconsistent with the new code and the new comment remains unchanged from the old comment, it indicates that the developer neglected to update the comment, resulting in an inconsistency within the 7 days preceding the new commit (termed Outdated data record). Otherwise, inconsistencies occurring earlier are termed Earlier Outdated data records. Data records where the new code aligns with the new comment are classified as Normal data records. Given that each of the N new commits can correspond to multiple data records due to discrepancies extending beyond a single method, this approach allows for comprehensive analysis.

To study the impact, we utilized odds ratios to assess the relationship between different types of data records and the likelihood of leading to bug-introducing commits. The odds ratio measures the probability of exposure when an event occurring compared to a non-exposed group. Here, the exposed group consists of Outdated and Earlier Outdated data records, while the non-exposed group comprises Normal data records. We calculated the odds of these groups leading to bug-introducing commits by comparing the number of records leading to such commits to those not leading to such commits within each group.

The general results on the number of records in exposed and non-exposed groups, and the occurrence or non-occurrence of the event for the 7-day window and 14-day window are demonstrated in Table \ref{general_7} and Table \ref{genral_14} respectively, For more detailed results in each repository, see Table \ref{detailed_7} and Table \ref{detailed_14} (RQ2 and RQ3). 
Based on the odds ratio analysis, inconsistent changes are approximately 1.5 times more likely to result in a bug-introducing commit than normal changes within a one-week window(RQ2), and 1.14 times more likely to do so within a two-week window.

\begin{finding}{Finding}
Inconsistent changes are approximately 1.52 times more likely to result in a bug-introducing commit compared to regular changes within a one-week period and 1.14 times more likely within a two-week period.
\end{finding}\bigskip

The detailed calculations show that inconsistent changes have a more significant impact on bug introduction within a 7-day window compared to a 14-day window. This indicates that bug-introducing commits are more likely to be influenced by recent inconsistent changes than those that occurred earlier.
It can be observed in Table \ref{two_weeks} that inconsistent changes in the week immediately preceding the target commits are more likely to result in bug-introducing commits compared to inconsistent changes occurring one week earlier.

\begin{finding}{Finding}
 Bug-introducing commits are more likely to be influenced by recent inconsistent changes than those that occurred earlier.
\end{finding}\bigskip

The results demonstrate that time affects the impact of code comment inconsistency on bug introduction. Over longer timeframes, the impact of inconsistency decreases, with the highest impact occurring shortly after the changes are made. Thus, inconsistent changes are most likely to introduce bugs in the near future rather than much later (RQ3).

\begin{table}
        \begin{tabular}{lll}\toprule
            &\textbf{Exposed group} & \textbf{Non-exposed group}\\\midrule
            Event happening & 2710 & 36672\\
            Event not happening& 2342 & 44907
            \\\bottomrule
        \end{tabular}
        \caption{General results for 7 days window}\label{general_7}
    \end{table} 

\begin{table}
        \begin{tabular}{lll}\toprule
            &\textbf{Exposed group} & \textbf{Non-exposed group}\\\midrule
            Event happening & 7021 & 102076\\
            Event not happening& 6654 & 110542
            \\\bottomrule
        \end{tabular}
        \caption{General results for 14 days window}\label{genral_14}
    \end{table}

\begin{table}
        \begin{tabular}{lll}\toprule
            &\textbf{Recent week} & \textbf{Earlier week}\\\midrule
            1 & 1.5126508 & 0.3026472693\\
            2 & 2.43915528 & 0.9414955803\\
            3 & 1.76433893 & 0.7395251739\\
            4 & 2.678552505 & 1.564742491\\
            5 & 0.7204445434 & 1.042315553\\
            6 & 1.502191465 & 2.844618366\\
            7 & 1.307834527 & 1.052682392\\
            8 & 1.239165345 & 1.16637326\\
            \textbf{Total} & \textbf{1.519541449} & \textbf{0.9460972249}
            \\\bottomrule
        \end{tabular}
        \caption{Comparaive results of different weeks}\label{two_weeks}
\end{table}

\begin{table*}[htb]
    \begin{tabularx}{\textwidth}{p{1cm}XXXXX}
    \toprule
       \textbf{Repo No.} &\textbf{Exposed group/Event happening} & \textbf{Non-exposed group/Event\newline happening} &\textbf{Exposed group/Event not happening} & \textbf{Non-exposed group/Event\newline not happening}  & \textbf{Odds ratio} \\\midrule
        1 & 640 & 4208 & 405 & 4028 & 1.5126508\\
        2 & 118 & 805 & 250 & 416 &2.43915528\\
        3 & 546 & 3398 & 608 & 6676 & 1.76433893\\
        4 & 730 & 3217 & 704 & 8310 & 2.678552505\\
        5 & 414 & 7331 & 437 & 5575 & 0.7204445434\\
        6 & 132 & 1156 & 90 & 1184 & 1.502191465\\
        7 & 1408 & 8654 & 704 & 5659 & 1.307834527\\
        8 & 1079 & 5414 & 340 & 2114 & 1.239165345\\
        \textbf{Total} & \textbf{5067} & \textbf{34183} & \textbf{3313} & \textbf{33962} & \textbf{1.519541449}
        \\\bottomrule
    \end{tabularx}
    \caption{Detailed results for 7 days window}\label{detailed_7}
\end{table*} 

\begin{table*}[htb]
    \begin{tabularx}{\textwidth}{p{1cm}XXXXX}
    \toprule
        \textbf{Repo No.} &\textbf{Exposed group/Event happening} & \textbf{Non-exposed group/Event\newline happening} &\textbf{Exposed group/Event not happening} & \textbf{Non-exposed group/Event\newline not happening}  & \textbf{Odds ratio} \\\midrule
        1 & 856 & 12594 & 769 & 8305 & 0.7340461627\\
        2 & 407 & 4145 & 109 & 1330 & 1.198105377\\
        3 & 1856 & 15901 & 1999 & 16494 & 0.9630896827\\
        4 & 2270 &9914 & 1748 & 15414 & 2.019067643\\
        5 & 1262 & 16644 & 909 & 10978 & 0.9157163974\\
        6 & 341 & 3951 & 114 & 2097 &1.587599409\\
        7 & 2862 &19139 & 1731 & 13455 & 1.162350271\\
        8 & 2958 &14562 & 1603 & 9286 & 1.176717737\\
        \textbf{Total} & \textbf{12812} & \textbf{96850} & \textbf{8982} & \textbf{77359} & \textbf{1.139344608}
        \\\bottomrule
    \end{tabularx}
    \caption{Detailed results for 14 days window}\label{detailed_14}
\end{table*} 

\begin{finding}{Finding}
  Over longer timeframes, the impact of inconsistency decreases, with the highest impact occurring shortly after the changes are made. Thus, inconsistent changes are most likely to introduce bugs in the near future rather than much later
\end{finding}\bigskip

\section{Discussion}

In the following, we discuss and compare the findings of studies by \mycite{tan2007icomment}, \mycite{khamis2010automatic}, and \mycite{ibrahim2012relationship} with our own results. We highlight how our detailed analysis, provided in previous sections, quantifies the impact of code comment inconsistency on bug introduction.

\mycite{tan2007icomment} analyzed lock-related comments on four projects and identified 60 inconsistencies, 27 of which were notably problematic. Some inconsistencies were acknowledged by developers. The researchers provided an example from the Mozilla project where an inconsistent comment caused bugs. While this study is valuable and pioneering in analyzing comments and their impact on bug introduction, its findings are limited due to the small sample size and focus solely on lock-related changes. Moreover, the study does not quantify the impact of these inconsistencies on bug introduction, unlike our odds ratio analysis. The study does not explore the relationship between code-comment inconsistencies and bugs through extensive analysis but rather relies on a limited number of inconsistencies identified.

\mycite{khamis2010automatic} examined the correlation between the quality of in-line documentation (comments) and the number of reported bug defects in software projects. While their findings suggest a relationship between Code/Comment Consistency metrics and a reduced number of reported bugs, it does not establish a direct causal link to bug introduction. Other unmeasured variables may also influence bug counts.
The correlation indicates a relationship between code comment inconsistency and bug frequency. However, the odds ratio analysis provides deeper insights into how inconsistency relates to bug introduction in this research. We specifically analyze the changes within 1-week and 2-week periods preceding each bug-introducing commit, rather than focusing on the overall number of bugs in the project.

\mycite{ibrahim2012relationship} refined traditional bug prediction models by incorporating comment update practices, distinguishing between consistent and inconsistent changes and bug fixes across three open-source projects (FreeBSD, PostgreSQL, and Eclipse), and they extracted the number of consistent and inconsistent changes and bug fixes from these repositories. They found that incorporating comment update practices into bug prediction models significantly improved the models' explanatory and predictive power. This improvement demonstrated an empirical link between comment update practices and future bugs. However, Potential inaccuracies in detecting changes to comments and code entities, along with the heuristic classification of changes as consistent or inconsistent without semantic consideration, may compromise the validity of their findings.

In this study, we examined 8 Apache repositories to enhance the generalizability of our findings. We employed Large Language Models (LLMs) to detect code comment inconsistencies, comparing this method with other established models in RQ1.  Our analysis utilized odds ratio analysis, offering straightforward and interpretable insights into the relationship between comment inconsistencies and bug-introducing likelihood. This method is well-suited for quantifying an association, especially in scenarios involving binary outcomes, such as bug presence or absence, which aligns with the focus of our study. We also examine the temporal evolution of comments and assess how time influences the ratio of code comment inconsistency's impact on bug introduction, a perspective that has not been explored in previous studies.

Our findings highlight the critical role that consistent and up-to-date code comments play in maintaining software quality. By investigating the relationship between code-comment inconsistency and bug introduction, we have provided evidence that inconsistencies between code and comments are a significant factor in the occurrence of bugs.

Our analysis shows that bug-prone commits are often associated with recent inconsistent changes. This emphasizes the importance of addressing inconsistencies as soon as possible to minimize their negative impact on software reliability. The temporal aspect of our findings, where the likelihood of bug introduction is highest shortly after an inconsistent change, further supports the need for timely maintenance of code comments.
The use of large language models, specifically GPT-3.5, in detecting code-comment inconsistencies demonstrates the potential of using LLM techniques in this domain. Our results indicate that GPT-3.5 outperforms traditional deep learning and machine learning approaches, providing more accurate and comprehensive identification of inconsistencies. This suggests that incorporating AI-driven tools into the software development lifecycle could enhance the detection and correction of inconsistencies, ultimately improving the quality of source codes.

\section{Threats to Validity}
In this section, we discuss the threats to the validity of our study and the measures taken to mitigate them. 

\subsection{Internal Validity}
One of the primary threats to the internal validity of our work is the accuracy of the Large Language Models (LLMs) used to detect code-comment inconsistencies. While LLMs like GPT-3.5 offer significant improvements over traditional methods, they are not infallible and may still produce false positives or negatives. To mitigate this, we performed fine-tuning and validation of the models on a subset of the cleared CUP dataset\cite{10172689} to enhance detection accuracy.

The impact of code-comment inconsistencies may vary over time. Our analysis considered the temporal evolution of comments, but there may still be unaccounted temporal factors influencing bug introduction. 
\subsection{External Validity}

Our study focuses on eight diverse Apache repositories, which may limit the generalizability of our findings to other software projects. While we selected repositories from different domains to enhance generalizability, caution should be exercised when applying these results to other contexts.

Identifying bug-introducing commits relies on historical data, which may be incomplete. Some bugs may have been fixed without proper documentation, leading to underreporting and affecting the accuracy of identification tools like GitCProc. To mitigate this, we used well-maintained and widely studied Apache repositories to ensure more reliable data. Additionally, we discarded commits that only changed documentation or tests from the non-bug-introducing set to enhance the accuracy of our analysis.

\section{Conclusion and Future Works} 

In this study, we first assessed GPT-3.5's ability to detect code-comment inconsistencies by comparing its performance to other state-of-the-art models, demonstrating its superior performance.

Then, using the superior model for code comment inconsistency detection (the LLM), we examined the impact of code-comment inconsistency on the introduction of bugs in software projects. By analyzing a comprehensive dataset of Java projects from the Apache repositories, we identified patterns indicating that inconsistencies between code and comments can contribute to the occurrence of bugs.

Our results reveal that bug-prone commits are more likely influenced by recent inconsistent changes rather than older ones. This suggests that the timeliness of addressing comment inconsistencies is critical in mitigating potential bug introduction. Additionally, our analysis demonstrated that the impact of these inconsistencies diminishes over time, with the highest likelihood of bug introduction occurring shortly after the inconsistent changes are made. This insight underscores the importance of continuous and timely maintenance of code comments to enhance software quality and reliability.

Based on the findings of this study, several avenues for future research are identified. Firstly, expanding the dataset to include a wider variety of programming languages and project types could enhance the generalizability of the results. 
Additionally, an extended study investigating the long-term impact of consistent code-comment practices on software maintainability and developer productivity would provide deeper insights into the practical benefits of the proposed methodologies.

\section{Acknowledgments}

\bibliographystyle{ACM-Reference-Format}
\bibliography{main}

\appendix

\end{document}